\documentclass[preprint,showpacs,preprintnumbers,amsmath,amssymb,nofootinbib]{revtex4-1}

\usepackage{graphicx}
\usepackage{dcolumn}
\usepackage{bm}

\newcommand{\beq}{\begin{equation}}
\newcommand{\eeq}{\end{equation}}
\newcommand{\beqy}{\begin{eqnarray}}
\newcommand{\eeqy}{\end{eqnarray}}
\newcommand{\beqyn}{\begin{eqnarray*}}
\newcommand{\eeqyn}{\end{eqnarray*}}
\newcommand{\nl}{\newline}
\newcommand{\nn}{\nonumber}

\newcommand{\bc}{\begin{center}}
\newcommand{\ec}{\end{center}}
\newcommand{\bmin}{\begin{minipage}}
\newcommand{\emin}{\end{minipage}}

\begin{document}

\title{A note on the implications of gauge invariance in QCD}

\author{Elliot Leader}
 \affiliation{Blackett laboratory \\Imperial College London \\ Prince Consort Road\\ London SW7 2AZ, UK}
\email{e.leader@imperial.ac.uk}
\author{Enrico Predazzi}
\affiliation{Dipartimento di Fisica Teorica,  Universit\`a di Torino,
\\ and INFN, Sezione di Torino, Italy}
\email{Predazzi@to.infn.it}
\date{\today}

\begin{abstract}
We compare and contrast the implications of gauge invariance for the structure of scattering amplitudes in QED and QCD. We derive the most general analogue for QCD of the famous QED rule that the scattering amplitude must be invariant if the polarization vectors $\epsilon^{{\mu}_j}(k_j)$ of any number of photons undergo the  replacement $ \epsilon^{{\mu}_j}(k_j) \rightarrow \epsilon^{{\mu}_j}(k_j) + c \, k_j^{\mu_j} $, where $c$ is an arbitrary constant.
\end{abstract}

\maketitle

\section{\label{Intro} Introduction}
In QED the amplitude for a general reaction involving $M$ incoming photons with momenta $k_1^{\mu_1}, k_2^{\mu_2}.......k_M^{\mu_M}$ and $N$ outgoing photons with momenta $\kappa_1^{\nu_1}, \kappa_2^{\nu_2}.........\kappa_N^{\nu_N}$ has the structure

\beq \label{AmpQED} \mathcal{A}= \epsilon^*_{\nu_1}(\kappa_1)......\epsilon^*_{\nu_N}(\kappa_N)\,M^{\nu_1....\, \nu_N}_{\mu_1....\, \mu_M} \, \epsilon^{\mu_1}(k_1)...... \epsilon^{\mu_M}(k_M)   \eeq
where the $\epsilon^{\mu}(k)$ are the photon polarization vectors.

It is very well known that ``as a consequence of gauge invariance"

\beq \label{GIQED} \kappa_{j,\, \nu_j}\,M^{\nu_1...\, \nu_j...\, \nu_N}_{\mu_1.........\, \mu_M}= 0 \quad \quad M^{\nu_1.........\, \nu_N}_{\mu_1...\, \mu_i...\, \mu_M}\,k_i^{\mu_i} = 0  \qquad (\textrm{any}\, i\,\in \, 1...M;\,j\, \in \, 1...N ) \eeq
irrespective of the value of $(\kappa_j)^2$ or $(k_i)^2$.

It is perhaps less well known that the  condition Eq.~(\ref{GIQED}) does \emph{not} hold for the QCD amplitude for  a reaction involving $M$ incoming and $N$ outgoing gluons.

The condition Eq.~(\ref{GIQED})is extremely important and useful:
\begin{itemize}
\item it simplifies the calculation of cross-sections from amplitudes;
\item it controls the allowed tensorial structure of amplitudes; a topical example is deep inelastic lepton-hadron scattering,
 where the interaction between the off-mass-shell photon of momentum $q$ and the hadron is described by the hadronic tensor $W^{\mu\nu}$, which must satisfy the condition $q_\mu \, W^{\mu\nu} = W^{\mu\nu}\, q_\nu =0 $;
 \item it provides a useful check on the correctness of calculations of scattering amplitudes.
    \end{itemize}

It is therefore of interest to know what relation in QCD is closest to the QED condition  Eq.~(\ref{GIQED}).

Some time ago, in the process of writing a textbook on gauge theories and particle physics \cite{Leader:1996hm} we discussed the difference
between the implications of gauge invariance for amplitudes in QED and QCD, explaining that the result Eq.~(\ref{GIQED}) in QED is
far from a \emph{trivial} consequence of electromagnetic gauge invariance in a \textit{quantum} field theory and showing why such a  relation could not be derived in QCD. We ended up deriving what we believed was the closest analogue in QCD to the QED relation Eq.~(\ref{GIQED}).
At the time, although we had not seen this result in the literature, we assumed it was well known to workers in the field.
Subsequently, as a result of questions raised by colleagues and comments made at workshops, we have come to realize
that, in fact, the result is not generally known, and that it ought, therefore, to be reported in the literature.
In this paper we present a brief discussion of the differences between QED and QCD and a derivation of our result for QCD. Our final result is presented as
a little Theorem at the end of Section IV.
\section{\label{sec1}QCD: a reminder}
$ G^a_{\mu \nu}$, the $SU(3)$ non-Abelian generalization of the standard QED field tensor $F^a_{\mu \nu}$ is given by

\begin{equation} \label{GQCD}
G^a_{\mu \nu}  ={ \partial}_{\mu}\, A^a_{\nu} - { \partial}_{\nu}\, A^a_{\mu} +
   g \,  f_{abc}\, A^b_{\mu}\, A^c_{\nu},
\end{equation}
where $A^a_{\mu}$ is the gluon vector potential, $ a = 1, 2,... 8$ is the octet colour label and $f_{abc}$ are the usual $SU(3)$ structure constants , i.e. the $SU(3)$ group generators $T_a$ obey the commutation relations
\begin{equation}
[T_a, T_b] = i \, f_{abc} \, T_c.
\end{equation}

The colour indices will be indifferently written as superscripts or subscripts. (Sometimes $-g$ is  used instead of $g$. This is of no relevance here given that all perturbative QCD calculations depend on $g^2$.)

The covariant derivative operator is defined as
\begin{equation} \label{CovDifOp}
\hat D_{\mu} \equiv \partial_{\mu} - i g T_a \, A^a_{\mu}.
\end{equation}

Acting on a given field which transforms according to a specific group representation, the $T_a$ are replaced by the corresponding representation matrices. As a consequence, acting on quark fields, Eq.~(\ref{CovDifOp}) becomes
\begin{equation}
( D_{\mu})_{ij} = {\delta}_{ij} \,  \partial_{\mu} - i g {t^a}_{ij} \, A^a_{\mu}
\end{equation}
where $t^a$, $a= 1, 2, ...8$ are $3 \times 3$ Hermitian matrices which for the triplet $SU(3)$ representation are $1/2$ the  Gell-Mann matrices $\lambda^a$.
Acting on the gluon fields, the $T_a$ are represented by the structure constants ${T_a}_{bc} \rightarrow -i f_{abc}$ and $\hat D_{\mu}$ is then represented by
\begin{equation} \label{CovDifG}
( D_{\mu})_{bc} = {\delta}_{bc} \,  \partial_{\mu} - i g f_{abc} \, A^a_{\mu}.
\end{equation}

\section{\label{sec2} The QCD currents : differences from QED}

The gauge invariant  Lagrangian density $\mathcal{L}$ is written as

\begin{equation} \label{QCDL}
\mathcal{L} = - \frac{1}{4} G^a_{\mu \nu} G^{ \mu \nu}_a \, + \, i \, \bar{\psi_i} \, \gamma_{\mu} \,
(D^{\mu})_{ij} \, {\psi}_j.
\end{equation}
Flavour summation, which is irrelevant for our discussion, is implied for the last term and we have, as usual, assumed massless quarks.

From Eq.~(\ref{QCDL}) one derives the equations of motion. For the gluon field we get
\begin{equation} \label{DG}
                     (D^{\mu})_{ab}  G^b_{\mu  \nu} = g J^a_{\nu}
\end{equation}
where the quark current is
\begin{equation}
                  J^a_{\nu} = {\bar {\psi}_i} {\gamma}_{\nu} {{t^a}_{ij}} {\psi}_j
\end{equation}

Now in QED the analogous electromagnetic current, the source of the photon field, is conserved, whereas the QCD current $J^a_{\nu}$ is not conserved in the usual sense, i.e. $\partial^{\mu} J^a_{\mu} \neq 0$. However, one finds that
\begin{equation}
                 (D^{\mu})_{ab} J^b_{\mu} =0 .
\end{equation}

It may appear puzzling that $J^a_{\mu}$ is not conserved since the theory is invariant under the $SU(3)_C$ group of colour gauge transformations. There \emph{are} indeed conserved Noether currents but they do not coincide with  $J^a_{\mu}$. The conserved Noether current is
\begin{equation}
                  \tilde J^a_{\mu} = J^a_{\mu} + f_{abc} G^b_{\mu  \nu} A_c^{\nu}.
\end{equation}

The interesting feature is that using Eq.(\ref{CovDifG}) in Eq.(\ref{DG}) one gets
\begin{equation} \label{divG}
                  {\partial}^{\mu} G^a_{\mu \nu} = g  \tilde J^a_{\nu}
\end{equation}

Notice, however, that, in contrast with QED, the current that appears in the above equation is not just a quark current but contains $G_{\mu \nu}$ itself, This stems from the fact that while the photon field is electrically neutral, the gluons have a colour charge. This, of course, is at the same time the richness and the complication of QCD compared with QED. Ultimately, this is the effect of the gluon self coupling which will be discussed in the next subsections.

Now in  a gauge theory one has to choose a definite gauge in which to work  and this gauge fixing is carried out in such a way as to preserve the invariance of the theory under some variant of the \emph{global} version of the original local gauge invariance. Concerning  the two currents we have introduced, the non-conserved $J^a_{\mu}$ and the conserved $\tilde J^a_{\mu}$, it is important to notice that neither is invariant under the QCD global transformations.

 With standard notations, under an infinitesmal SU(3) non Abelian gauge transformation, the various QCD fields transform as
\begin{equation}
                     {\delta{\psi_j}} = -i \, {t^b}_{jk} \, \psi_k \, \theta_b,
\end{equation}
\begin{equation}
                    {\delta{\bar \psi_j}} = i \,{\bar \psi_k} \, {t^b}_{jk} \,  \theta_b,
\end{equation}
\begin{equation}
                    \delta A^a_{\mu} =  f_{abc} A^c_{\mu} \,  \theta_b,
\end{equation}
\begin{equation}
                   \delta G^a_{\mu \nu} = f_{abc} \, G^a_{\mu \nu}  \,  \theta_b.
\end{equation}

Using these transformations, we see that both the currents $J^a$ and $\tilde J^a_{\mu}$ are not gauge invariant. For example, for $J^a$ we obtain
           \beq    \label{GTJa}    \delta J^a_{\mu} =  f_{abc} \, J^c_{\mu}  \,  \theta_b.
\end{equation}
and an analogous equation for $\tilde J^a_{\mu}$, where $\theta_b$ is a constant phase.

In contrast, recall that in the Abelian case of QED. the electromagnetic current \emph{is} invariant under gauge transformations {\it i.e.}
\begin{equation} \label{GTemJ}
                   \delta J^{em}_{\mu} =  0.
\end{equation}

Now it can be shown that the generators of any symmetry transformation are the conserved {\it charges} associated with the Noether currents. Thus, in the em case
the generator is
\begin{equation}
                  \hat Q = \int d^3 x  \, J^{em}_0(\bm{x},t)
\end{equation}
and, if the change induced in any function $F(x)$ of the field operators by an infinitesimal em gauge transformation is
\begin{equation}
                   F(x) \rightarrow  F(x) + \delta F(x),
\end{equation}
then
\begin{equation}
                  \delta F(x) \propto [\hat Q, F(x)].
\end{equation}

Owing to the gauge invariance of $J^{em}_{\mu}$, we then have
\begin{equation} \label{JJemCom}
                    \int d^3 x  \,[J^{em}_0(\bm{x},t),  J^{em}_{\mu}(y)] \, = \, 0.
\end{equation}

In the QCD case, there are eight  generators $\hat Q^a$ associated with the eight Noether currents $ \tilde J^a_{\mu}$ via
\begin{equation}
                  \hat Q^a = \int d^3 x  \, \tilde J^a_0(\bm{x},t)
\end{equation}
and
in contrast to QED, since neither $J^a_{\mu}$ nor $ \tilde J^a_{\mu}$ are invariant under gauge transformations, the analogue of Eq.~(\ref{JJemCom}) does not hold and
\begin{equation} \label{JJqcdCom}
        \int d^3 x  \,[\tilde J^a_0(\bm{x},t),  J^b_{\mu}(y) \,\, \textrm{or}\,\, \tilde J^b_{\mu}(y)] \, \neq \, 0.
\end{equation}

\section{ \label{sec3} Consequences of gauge invariance for matrix elements}

We shall now study the implications of the above on matrix elements in QED and QCD.

\subsection{On the substitution $\epsilon^{\mu}(k) \rightarrow \epsilon^{\mu}(k) + c k^{\mu}$ for the photon polarization vector in QED}

Let us consider Compton scattering as a typical QED reaction
\begin{equation}
         \gamma (k) + e(p) \rightarrow  \gamma (k') + e(p')
\end{equation}
whose scattering amplitude will be of the form
\begin{equation} \label{QEDME}
        (S-1)_{fi} \, = \, \epsilon ^{\mu \star}(k') M_{\mu \, \nu} \, \epsilon ^{\nu}(k)
\end{equation}

It is often stated that, as a trivial consequence of gauge invariance, Eq.~(\ref{QEDME}) is unchanged when the photon polarization vector $\epsilon^{\mu}(k)$ is replaced by
\begin{equation} \label{epsilonGT}
\epsilon^{\mu}(k) \rightarrow \epsilon^{\mu}(k) + c k^{\mu}
\end{equation}
where $c$ is an arbitrary constant,  leading to the
fundamental relations
\begin{equation}\label{GInMEem}
          {k'}^{\mu}\, M_{{\mu}{\nu}} \, = \, M_{{\mu}{\nu}} \, k^{\nu} \, = \, 0,
\end{equation}
and strictly  analogous relations hold for QED processes involving any number of external photons (see Eq.~(\ref{GIQED}) ).

In fact relation Eq.~(\ref{GInMEem}) is far from a \emph{trivial} consequence of gauge invariance.

In classical electrodynamics, the invariance under $\epsilon^{\mu}(k) \rightarrow \epsilon^{\mu}(k) + c k^{\mu}$ follows from the gauge invariance of the theory under the local gauge transformation
\begin{equation} \label{qedgin}
           A_{\mu}(x) \rightarrow  A_{\mu}(x) \, + \, \frac{1}{e}\, \partial_{\mu} \Lambda(x)
\end{equation}
where $\Lambda(x)$ is an arbitrary function of $x$. Eq.~(\ref{epsilonGT}) is the Fourier transform of Eq.~(\ref{qedgin}) when  $ A_{\mu}(x)$ is a plane wave. This is fine in classical electrodynamics but not in QED on two counts. First, the replacement  Eq.~(\ref{epsilonGT}) follows from Eq.~(\ref{qedgin}) only if $\Lambda(x)$ is not just an arbitrary function of $x$ but a scalar quantized field linear in the photon creation and annihilation operators. Secondly, we have to make a choice of gauge and once this is done, the theory is no longer invariant under \emph{local} gauge transformations.

We can, however, always make a gauge choice so that {\it global} gauge invariance is preserved. In this case, the conditions Eq.~(\ref{GIQED}) satisfied by the QED matrix elements follows from three properties:

(i) The global gauge invariance implies the existence of a conserved Noether current $J^{em}_{\mu}(x)$;

(ii) $A_{\mu}(x)$ couples to this conserved current in the QED Lagrangian;

(iii) The current itself is gauge invariant.

To prove that the above statements {\it i.e.} to prove that (i)-(iii) imply invariance under the replacement Eq.~(\ref{epsilonGT}), we resort to the LSZ reduction formalism (see ref \cite{Bjorken:1965dk}) applied to  Compton scattering  as an example. We shall derive the first of the relations in Eq.~(\ref{GInMEem}) in some detail. \nl

The amplitude $M_{\mu \,\nu}$ in Eq.~(\ref{QEDME}) is given by
\begin{equation} \label{LSZQED}
            M_{\mu \,\nu } = (-ie)^2 \, \int d^4 x \, d^4 y \exp{(ik'y-ikx)} \langle \, p'\, |\, T \, [J^{em}_{\mu}(y) \,   J^{em}_{\nu}(x)]\, |\, p \, \rangle
\end{equation}
where $T$ is the time ordering operator, and where
\begin{equation} \label{BoxA}
         \Box A_{\mu}(x)  = e  J^{em}_{\mu}(x),
\end{equation}
which is a consequence of (ii), has been used.

From Eq.~(\ref{LSZQED}) we get
\begin{equation}
          { k'}^{\mu} \, M_{\mu \, \nu} = (1/i) (-ie)^2 \, \int d^4 x \, d^4 y  \left[\frac{\partial}{\partial y_{\mu}} \,\exp{(ik'y)} \right] \exp{(-ikx)}\, \langle \,p'\, |\, T \, [J^{em}_{\mu}(y) \,   J^{em}_{\nu}(x)]\, |\, p \, \rangle.
\end{equation}

Integrating by parts and discarding the surface terms we obtain
\begin{equation} \label{k'diff}
            k'^{\mu} \, M_{\mu\,\nu} = i \, (-ie)^2 \, \int {d^4}x \, {d^4}y \exp{(ik'y-ikx)} \langle \, p' \,| \frac{\partial}{\partial y_{\mu}} \, T  \, [J^{em}_{\mu}(y) \,   J^{em}_{\nu}(x)] |\, p \, \rangle.
\end{equation}

The spatial derivatives are not influenced by the time ordering so that
\begin{equation}
             \frac{\partial}{\partial y_i} \,  T \, [J^{em}_{\mu}(y) \,  J^{em}_{\nu}(x)] \, = \, T \, \left[\frac{\partial J^{em}_{\mu}(y)}{\partial y_i} \,  J^{em}_{\nu}(x) \right]
\end{equation}
while the time derivative gives
\beqy             \frac{\partial}{\partial y_0} \,  T \, [J^{em}_{0}(y) \,  J^{em}_{\nu}(x)] & = &\, \frac{\partial}{\partial y_0}\, [ \theta(y_0 -x_0) \, J^{em}_{0}(y) \,  J^{em}_{\nu}(x)
 + \theta(x_0-y_0) J^{em}_{\nu}(x) \, J^{em}_{0} (y)] \nn \\
    & & \kern-6em + \, \delta(y_0-x_0) \, [ J^{em}_{0}(y) \,  J^{em}_{\nu}(x) \, - \, J^{em}_{\nu}(x) \, J^{em}_{0}(y)]  \, + \, T \left[\frac{\partial J^{em}_{\mu}(y)}{\partial y_0} \,  J^{em}_{\nu}(x) \right].
\eeqy

Adding together the last two equations, we get
\beqy \label{diffT}
\frac{\partial}{\partial y_{\mu}}\, T \,[J^{em}_{\mu}(y) \,   J^{em}_{\nu}(x)] \,& =& \, \delta(y_0-x_0) \,  [ J^{em}_{0}(y),J^{em}_{\nu}(x)] \nn \\
 &+& \, T \left[\frac{\partial J^{em}_{\mu}(y)} {\partial y_{\mu}} \,  J^{em}_{\nu}(x) \right].
\eeqy

The last term of Eq.~(\ref{diffT}) is zero for the conserved electromagnetic current. The other term is an equal-time commutator which can be evaluated explicitly taking $J^{em}_{\mu}$ of the form $\bar \psi {\gamma_{\mu}} \psi$, yielding
\begin{equation} \label{miracle}
                 [ J^{em}_{\mu}(y), J^{em}_{\nu}(x)]_{x_0 = y_0} \, = \, \bar \psi(x) [\gamma_{\mu} \gamma_0 \gamma_{\nu} - \gamma_{\nu} \gamma_0 \gamma_{\mu}] \psi(x)  \delta^3(\bm{x} - \bm{y}).
\end{equation}
It follows  that the equal time commutator in Eq.~(\ref{diffT}) vanishes \textit{i.e.}
\begin{equation} \label{ECT}
                   [ J^{em}_{0}(y) \, ,  J^{em}_{\nu}(x)]_{x_0 = y_0} \, = \, 0
\end{equation}
and thus the entire RHS of Eq.~(\ref{diffT}) vanishes. Substituting this result into  Eq.~(\ref{k'diff}) we obtain the desired result
  \beq \label{resQED} { k'}^{\mu} \, M_{\mu \, \nu} =0 .\eeq
   With inessential practical complications, this  generalizes to reactions with an arbitrary number of external photons, leading to the result Eq.~(\ref{GIQED}).

The seemingly miraculous vanishing of the equal time commutator in Eq.~(\ref{miracle}) is actually due to the gauge invariance of $J^{em}_{\mu}$ and is consistent with Eq.~(\ref{JJemCom}). As we shall see shortly, this is exactly the point where the analogy with QED will break down in QCD.

\subsection{On the substitution $\epsilon^{\mu}(k) \rightarrow \epsilon^{\mu}(k) + c k^{\mu}$ for the gluon polarization vector in QCD}

Let us now turn to the QCD analogue of Compton scattering \textit{i.e.} elastic gluon-quark scattering
\begin{equation}
                G{^b}(k) \, + \, q(p) \rightarrow   G{^a}(k') \, + \, q(p').
\end{equation}
where for clarity we will suppress the colour labels of the quarks.
The analogue of Eq.~(\ref{LSZQED}) would now read
\begin{equation} \label{MQCD}
            {M^{ab}}_{{\mu}{\nu}} = (-ig)^2 \, \int {d^4}x \,{d^4}y \exp{(ik'y-ikx)} \langle \, p'\, | \,T \,[{\hat J^{a}}_{\mu}(y) \,   {\hat J^{b}}_{\nu}(x)] \,|\, p \, \rangle
\end{equation}
where the current $\hat J^{a}_{\mu}(x)$ ``drives" the gluon field \textit{i.e.} the analogue of Eq.~(\ref{BoxA}) is
\begin{equation}
         \Box A^a_{\mu}(x)  = g  {\hat J^{a}}_{\mu}(x).
\end{equation}

From Eqs.(\ref{GQCD}, \ref{divG}) one finds
\begin{equation} \label{Jtilde}
          {\hat J^{a}}_{\mu}(x) \, = \, \tilde J^{a}_{\mu}(x) \, - \, f_{abc} \partial^{\nu} ({A^b}_{\nu}{A^c}_{\mu}).
\end{equation}

The current $\hat J^{a}_{\mu}(x)$ is conserved because a) the Noether current $\tilde J^{a}_{\mu}(x)$  is and b) because the last term in Eq.~(\ref{Jtilde}) being antisymmetric in the colour indices vanishes when acted on by $\partial^{\mu} $.

However, the current  $\hat J^{a}_{\mu}(x)$, like $J^a_\mu $ and  $ \tilde J^{a}_{\mu}(x)$   discussed earlier, is \emph{not}  gauge invariant. As a consequence, if, starting from Eq.~(\ref{MQCD}), one repeats the steps leading from Eq.~(\ref{LSZQED}) to  Eq.~(\ref{diffT})  there is now no reason that the equal time commutator in the analogue of Eq.~(\ref{diffT})    should vanish \textit{i.e.}   the QCD analogue of Eq.~(\ref{ECT}) is now
\begin{equation}
                 [\hat{J^{a}}_{0}(y)\, , \, \hat{J^{b}}_{\nu}(x)]_{x_0 = y_0} \, \neq  0.
\end{equation}

Thus we now find that
 \begin{equation}
          {k'}^{\mu}\, M^{ab}_{\mu \nu} \neq 0
\end{equation}
and similarly that $ \, M^{ab}_{\mu \nu} \, k^{\nu} \neq 0$.

However, it is also possible to approach the problem  from a slightly different angle and write for the scattering amplitude, instead of Eq.(\ref{QEDME}),
\begin{equation}
        (S-1)_{fi} \, = \, {{\epsilon}^{\mu}}^{\star}(k') M^{ab}_{\mu}
\end{equation}
where, for QCD
\begin{equation}
                M^{ab}_{\mu} = (-ig) \int d^4y \,\exp{(ik'y)} \langle \, p'\, |\,\hat J^{a}_{\mu}(y) \,|\,  p \, ; k, b \, \rangle .
\end{equation}
A simple integration by parts  now shows that
\begin{equation} \label{k'M}
{k'}^{\mu}\, M^{ab}_{\mu} \, = \, 0.
\end{equation}
purely as a consequence of the conservation of $\hat J^{a}_{\mu}$.

Since, however,
\begin{equation}
                M^{ab}_{\mu} = \, M^{ab}_{\mu \nu} \,\epsilon^\nu(k)
\end{equation}
Eq.~(\ref{k'M}) implies that
\begin{equation} \label{QCDrest}
                 k'^{\mu} M^{ab}_{\mu \nu}\, \epsilon^{\nu}(k) \, = \, 0,
\end{equation}
which is, of course, a weaker result than Eq.~(\ref{resQED}).

However, $\epsilon^{\nu}(k)$ is, apart from the condition $k \cdot \epsilon = 0$ and its normalization which is irrelevant,  an \emph{arbitrary} 4-vector. So Eq.~(\ref{QCDrest}) must hold also if $\epsilon^{\nu}(k)$ is replaced by $k^{\nu}$ \emph{provided that} $k^2=0$ \textit{i.e.}
\beq \label{ImpQCDres} k'^{\mu} M^{ab}_{\mu \nu}\, k^\nu \, = \, 0, \quad \textrm{for}  \quad   k^2=0. \eeq

It is straightforward to generalize this result to a reaction involving $M$ incoming gluons with momentum $k_j^{\mu_j}$ and colour $b_j$ and $N$ outgoing gluons, momentum $\kappa_i^{\nu_i}$ and colour $a_i$.

\textbf{Theorem}: If the QCD amplitude, analogously to the QED amplitude Eq.~(\ref{AmpQED}), is written
\begin{equation}\label{GIQCD}
\mathcal{A}_{QCD} = \epsilon_{\nu_1}^{\star}(\kappa_1) .......\epsilon_{\nu_N}^{\star}(\kappa_N)\, M_{a_1 .....a_N \, ; \,\, b_1 .....b_M}^{\nu_1 ..... \nu_N \,; \,\, \mu_1 .....\mu_M} \, \epsilon_{\mu_1}(k_1) .......  \epsilon_{\mu_M}(k_M)
 \eeq
 then one gets zero if  \emph{any number}, $\geq 1$, of the $\epsilon_{\mu_j}(k_j)$ and/or $\epsilon^*_{\nu_i}(k_i)$ are replaced by $k_{j\, , \, \mu_j}$ and/or  $\kappa_{i\, , \, \nu_i} $ respectively, {\emph{provided} that all these $k_j$'s and $\kappa_i$'s, with the exception of \emph{at most one of them}, satisfy $k_j^2 =0$ and  $\kappa_i^2=0$.

 \section{Conclusions}

 The powerful condition Eq.~(\ref{GIQED}),  which is valid in QED and which is extremely useful in calculating cross-sections, and determining the tensorial structure of amplitudes, unfortunately does not hold in QCD. The most general analogue of this,  valid in QCD,  is our result Eq.~(\ref{GIQCD}). Although not as useful as Eq.~(\ref{GIQED})  it is nonetheless helpful as a test of the correctness of amplitudes calculated in perturbative QCD.



\bibliography{Elliot_General}

\end{document}